\documentclass[11pt,preprint]{aastex}
\usepackage{multicol}

\newcommand{\kms}{\mbox{km\thinspace s$^{-1}$}} 

\makeatletter

\newenvironment{figurehere}
  {\def\@captype{figure}}
  {}
  \makeatother
\defcitealias{dull97}{D97}

\slugcomment{}
\shorttitle{The Dynamics of M15}
\shortauthors{Dull et al.}

\begin{document}

\title{ADDENDUM: ``THE DYNAMICS OF M15: OBSERVATIONS OF THE VELOCITY DISPERSION
       PROFILE AND FOKKER-PLANCK MODELS'' (ApJ, 481, 267 [1997])}

\author{J.D. Dull\altaffilmark{1}, H.N. Cohn\altaffilmark{2},
P.M. Lugger\altaffilmark{2}, B.W. Murphy\altaffilmark{3},
P.O. Seitzer\altaffilmark{4}, P.J. Callanan\altaffilmark{5}, 
R.G.M. Rutten\altaffilmark{6}, P.A. Charles\altaffilmark{7}}

\altaffiltext{1}{Department of Physics, Albertson College of Idaho,
2112 Cleveland Blvd., Caldwell, ID 83605}

\altaffiltext{2}{Department of Astronomy, Indiana University, 
727 E. Third Street, Bloomington, IN  47405}

\altaffiltext{3}{Department of Physics and Astronomy,
Butler University, 4600 Sunset Ave., Indianapolis, IN 46208}

\altaffiltext{4}{Department of Astronomy, University of Michigan,
830 Dennison Bldg., Ann Arbor, MI  48109}

\altaffiltext{5}{Department of Physics, National University of Ireland,
University College, Cork, Ireland}

\altaffiltext{6}{Isaac Newton Group of Telescopes, Apartado de correos 321,
E-38700 Santa Cruz de la Palma, Tenerife, Spain}

\altaffiltext{7}{Department of Physics and Astronomy, University of Southampton,
Southhampton SO17~1BJ, UK}

\email{cohn@astro.indiana.edu, bmurphy@butler.edu}

~~ 
\begin{multicols}{2}

It has recently come to our attention that there are axis scale errors
in three of the figures presented in \citet[][hereafter D97]{dull97}.
This paper presented Fokker-Planck models for the collapsed-core
globular cluster M15 that include a dense, centrally concentrated
population of neutron stars and massive white dwarfs.  These models do
not include a central black hole.  Figure 12 of D97, which presents
the predicted mass-to-light profile, is of particular interest, since
it was used by \citet{gers02a} as an input to their Jeans equation
analysis of the HST-STIS velocity measurements reported by
\citet{mare02}.  Based on the original, incorrect version of Figure
12, \citet{gers02a} concluded that the D97 models can only fit the new
data with the addition of an intermediate-mass black hole.  However,
this is counter to our previous finding, shown in Figure 6 of
\citetalias{dull97}, that the Fokker-Planck models predict the sort of
moderately rising velocity dispersion profile that \citet{gers02a}
infer from the new data.  \citet{baum02} have independently noted this
apparent inconsistency.

We appreciate the thoughtful cooperation of Roeland van der Marel in
resolving this issue.  Using our corrected version of Figure 12 (see
below), \citet{gers02b} now find that the velocity dispersion profile
that they infer from the \citetalias{dull97} mass-to-light ratio
profile is entirely consistent with the velocity dispersion profile
presented in Figure 6 of \citetalias{dull97}.  \citet{gers02b} further
find that there is no statistically significant difference between the fit
to the \citet{mare02} velocity measurements provided by the
\citetalias{dull97} intermediate-phase model and that provided by
their model, which supplements this \citetalias{dull97} model with a
$1.7_{-1.7}^{+2.7}\times10^3~M_\odot$ black hole.  Thus, the choice
between models with and without black holes will require additional
model predictions and observational tests.

We present corrected versions of Figures 9, 10, and 12 of
\citetalias{dull97}.  We take responsibility for the errors in the
original versions of these figures and regret any confusion that these
may have caused.  We also present an expanded version of Figure 6,
which extends the radial scale to both smaller and larger values, in
order to show the full run of the velocity dispersion profile.  The
profile of the intermediate-phase model of \citetalias{dull97} is
in good agreement with the HST-STIS velocity dispersion profile
presented by \citet{gers02a}.  In particular, the central value of
$\sim14~\kms$, predicted by this model, nicely coincides with their
findings.

We note that three independent studies have now demonstrated that
there is a dense, central concentration of dark mass in M15, by use of
three alternative methods: Fokker-Planck simulations
\citepalias{dull97}, GRAPE-6 simulations \citep{baum02}, and Jeans
equation modeling \citep{gers02a, gers02b}.  The dark mass is proposed
to consist of neutron stars and massive white dwarfs, in the former
two studies, versus a central black hole in the latter.  Irrespective
of these different interpretations of the nature of the dark mass, its
presence now appears to be well established on dynamical grounds.

\end{multicols}
~~\bigskip

\setlength{\columnsep}{50pt}

\newpage

\setlength{\columnsep}{35pt}
\begin{multicols}{2}

\begin{figurehere}
\centering\includegraphics[width=3.0in]{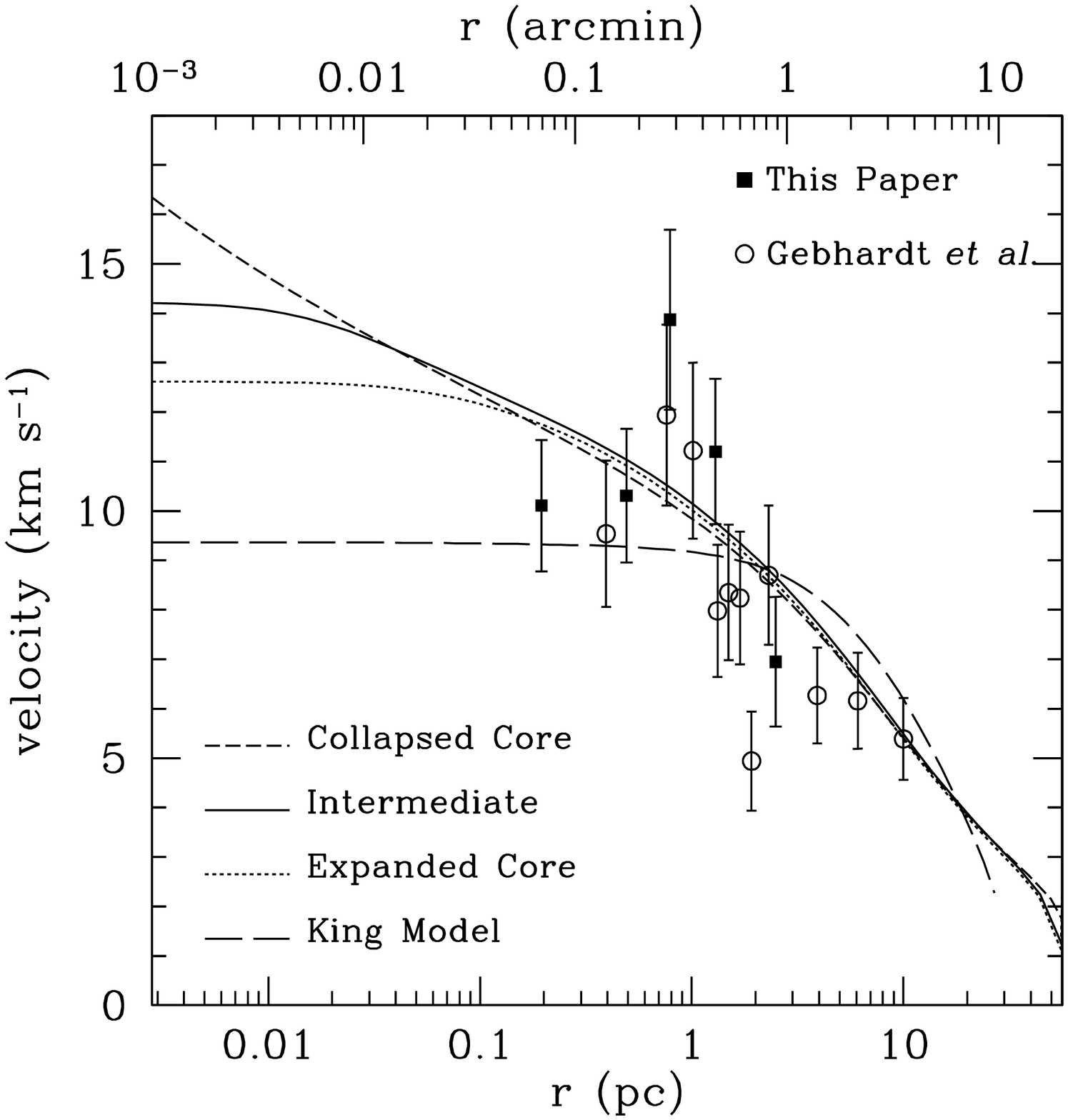}
{Fig.~6.--- Velocity dispersion versus radius for stars in M15\@.  The
filled squares are the data from \citetalias{dull97} and the open
circles represent the \citet{gebh94} data.  The radial scale has been
extended both inwards and outwards, relative to the original version
of this figure, to allow comparison with the much broader radial
coverage that is now available.}
\end{figurehere}

\begin{figurehere}
\centering\includegraphics[width=3.0in]{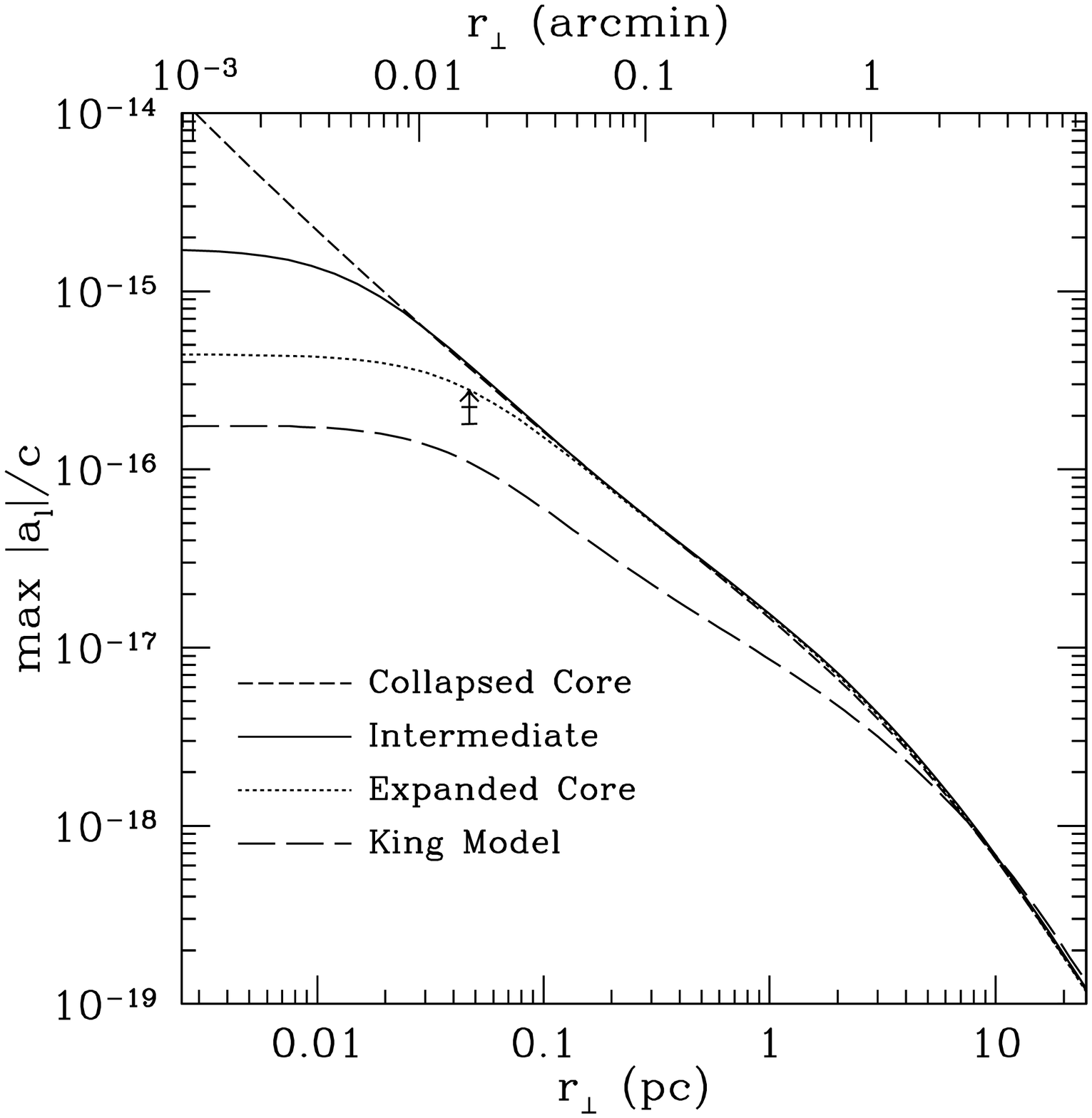}
{Fig.~9.--- The LOS acceleration as produced by the mean-field
gravitational potential of the best-fitting models.  The data symbol
indicates the minimum LOS acceleration needed to account for the
negative period derivatives of PSR~2127+11A (lower horizontal line)
and PSR~2127+11D (upper horizontal line); the pulsars lie at nearly
the same distance from the cluster center.  The radial scale of this
figure has been corrected from the original version, increasing every
radius by a factor of 2.82.  The top and bottom axes have been
interchanged, for consistency with the other figures.}
\end{figurehere}
\end{multicols}

\newpage

\begin{multicols*}{2}

\begin{figurehere}
\centering\includegraphics[width=3.0in]{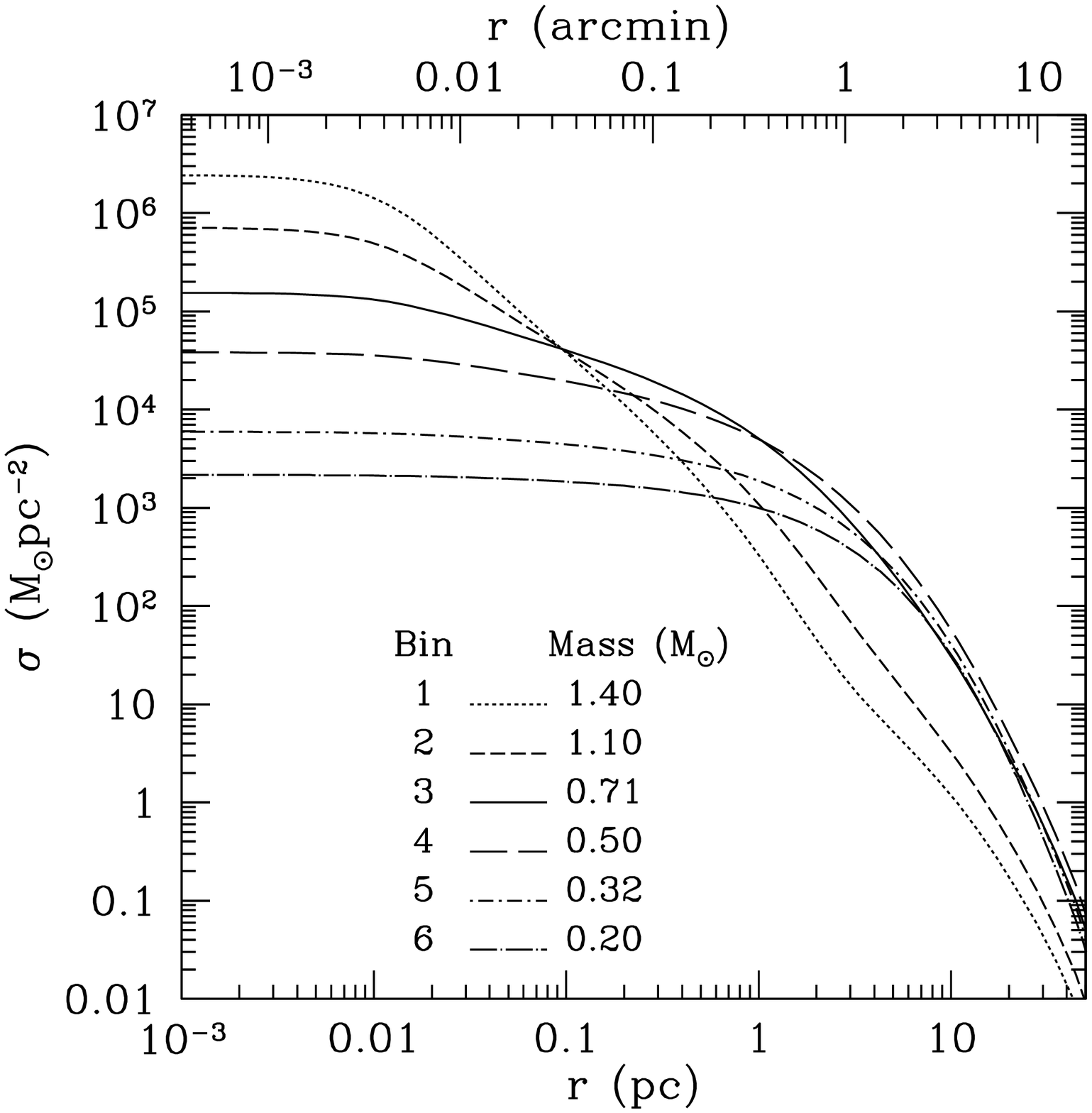}
{Fig.~10.--- Surface density profiles for the six mass groups of the
intermediate-phase Fokker-Planck model.  At small radius, the curves
are arrayed in order of increasing stellar mass, illustrating the
effect of mass segregation.  Note that groups 1 and 2 (neutron stars
and heavy white dwarfs) are strongly centrally concentrated relative
to the luminous stars.  The vertical scale of this figure has been
corrected from the original version, increasing all of the densities
by a factor of 2.53.}
\end{figurehere}

\begin{figurehere}
\centering\includegraphics[width=3.0in]{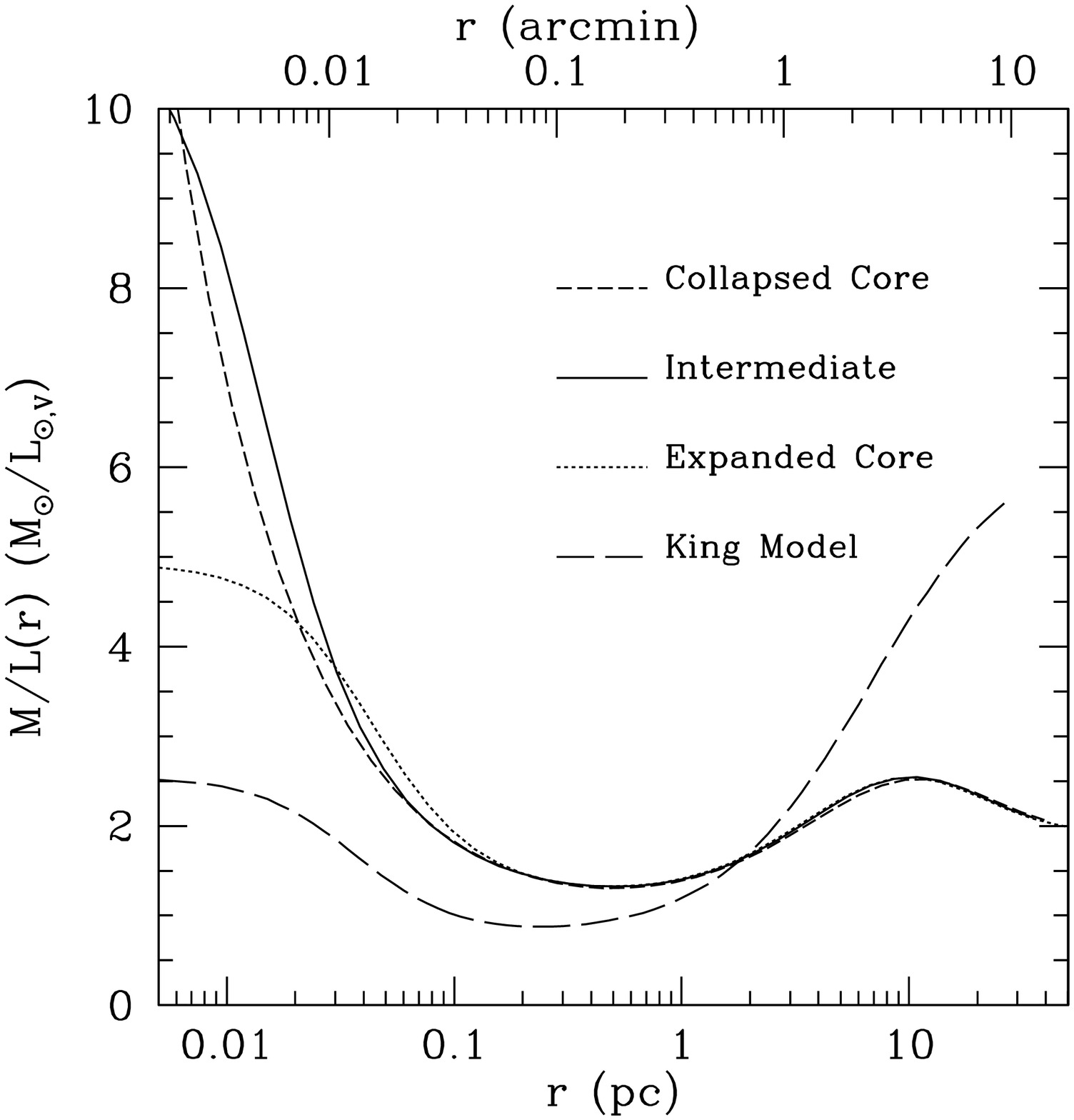}
{Fig.~12.--- Mass-to-light ratio profile for the three best-fitting
Fokker-Planck models and the King model.  $M/L$ is calculated as the
ratio of surface mass density to surface brightness.  The strong rise in
$M/L$ at small radius is due to the centrally concentrated populations
of nonluminous remnants seen in Figs.~10 and 11.  The radial scale of this
figure has been corrected from the original version, increasing every
radius by a factor of 2.82.  The top and bottom axes have been
interchanged, for consistency with the other figures.}
\end{figurehere}
\end{multicols*}

\end{document}